\documentclass[letterpaper]{article} 
\usepackage{aaai2026}  
\usepackage{times}  
\usepackage{helvet}  
\usepackage{courier}  
\usepackage[hyphens]{url}  
\usepackage{graphicx} 
\usepackage{natbib}  
\usepackage{caption}
\usepackage{framed}
\usepackage{amsmath}
\usepackage{amssymb}

\usepackage{algorithm}
\usepackage{algorithmic}

\usepackage{newfloat}
\usepackage{listings}
\DeclareCaptionStyle{ruled}{labelfont=normalfont,labelsep=colon,strut=off} 
\floatstyle{ruled}
\newfloat{listing}{tb}{lst}{}
\floatname{listing}{Listing}
\title{Path to Intelligence: Measuring Similarity between Human Brain and Large Language Model Beyond Language Task}
\author {
    Doai Ngo\textsuperscript{\rm 1},
    Mingxuan Sun\textsuperscript{\rm 2},
    Zhengji Zhang\textsuperscript{\rm 3},
    Mark Schnitzer \textsuperscript{\rm 3},
    Ashwin G Ramayya\textsuperscript{\rm 4},
    Zhe Zhao \textsuperscript{\rm 1}
}
\affiliations {
    \textsuperscript{\rm 1}Department of Computer Science, University of California, Davis\\
    \textsuperscript{\rm 2}Department of Statistics, University of California, Davis\\
    \textsuperscript{\rm 3}Department of Applied Physics, Stanford University\\
    \textsuperscript{\rm 4}Department of Neurosurgery, Stanford University\\
    dxngo@ucdavis.edu, 
    mxsun@ucdavis.edu, 
    zhjzhang@stanford.edu,
    aramayya@stanford.edu,
    mschnitz@stanford.edu,
    zao@ucdavis.edu
}

\usepackage{bibentry}

\begin{document}

\maketitle


\begin{abstract}
    Large language models (LLMs) have demonstrated human-like abilities in language-based tasks. While language is a defining feature of human intelligence, it emerges from more fundamental neurophysical processes rather than constituting the basis of intelligence itself. In this work, we study the similarity between LLM internal states and human brain activity in a sensory-motor task rooted in anticipatory and visuospatial behavior. These abilities are essential for cognitive performance that constitute human intelligence. We translate the sensory-motor task into natural language in order to replicate the process for LLMs. We extract hidden states from pre-trained LLMs at key time steps and compare them to human intracranial EEG signals. Our results reveal that LLM-derived reactions can be linearly mapped onto human neural activity. These findings suggest that LLMs, with a simple natural language translation to make them understand temporal-relevant tasks, can approximate human neurophysical behavior in experiments involving sensory stimulants. In all, our contribution is two-fold: (1) We demonstrate similarity between LLM and human brain activity beyond language-based tasks. (2) We demonstrate that with such similarity, LLMs could help us understand human brains by enabling us to study topics in neuroscience that are otherwise challenging to tackle.

\end{abstract}


\maketitle

\section{Introduction}
    Large Language Models (LLMs) have demonstrated impressive generalization capabilities across a wide range of language-based tasks \cite{xu2023expertprompting, shin2024caus}. These advances have inspired a wave of research exploring whether LLMs may serve as computational analogs of human cognition, particularly in language processing domains \cite{wang2023aligninglargelanguagemodels}. However, LLMs are trained solely on languages while human nervous systems encode and integrate information acquired from the environment through sensory inputs \cite{mattson_superior_2014}. Language-based tasks represent only a subset of the full range of human cognitive abilities, which also include visuospatial reasoning and motor control \cite{intelligence_theory}. 
    
    Motivated by this difference between human and artificial intelligence systems and the gap in this LLM alignment in general tasks, we investigate whether there exists any shared dynamics between LLMs and human brain activities in non-linguistic tasks. By doing so, we hope to compare and contrast artificial vs. biological intelligent systems and to gain insights about the elements that constitute intelligence. We focus specifically on anticipatory behavior in a sensory-motor task that engages visual processing  — core processes involved in everyday human cognitive functions and a building block for functions involving higher level of intelligence \cite{intelligence_theory}.
    
\begin{figure*}
        \centering
        \includegraphics[width=1.0\textwidth]{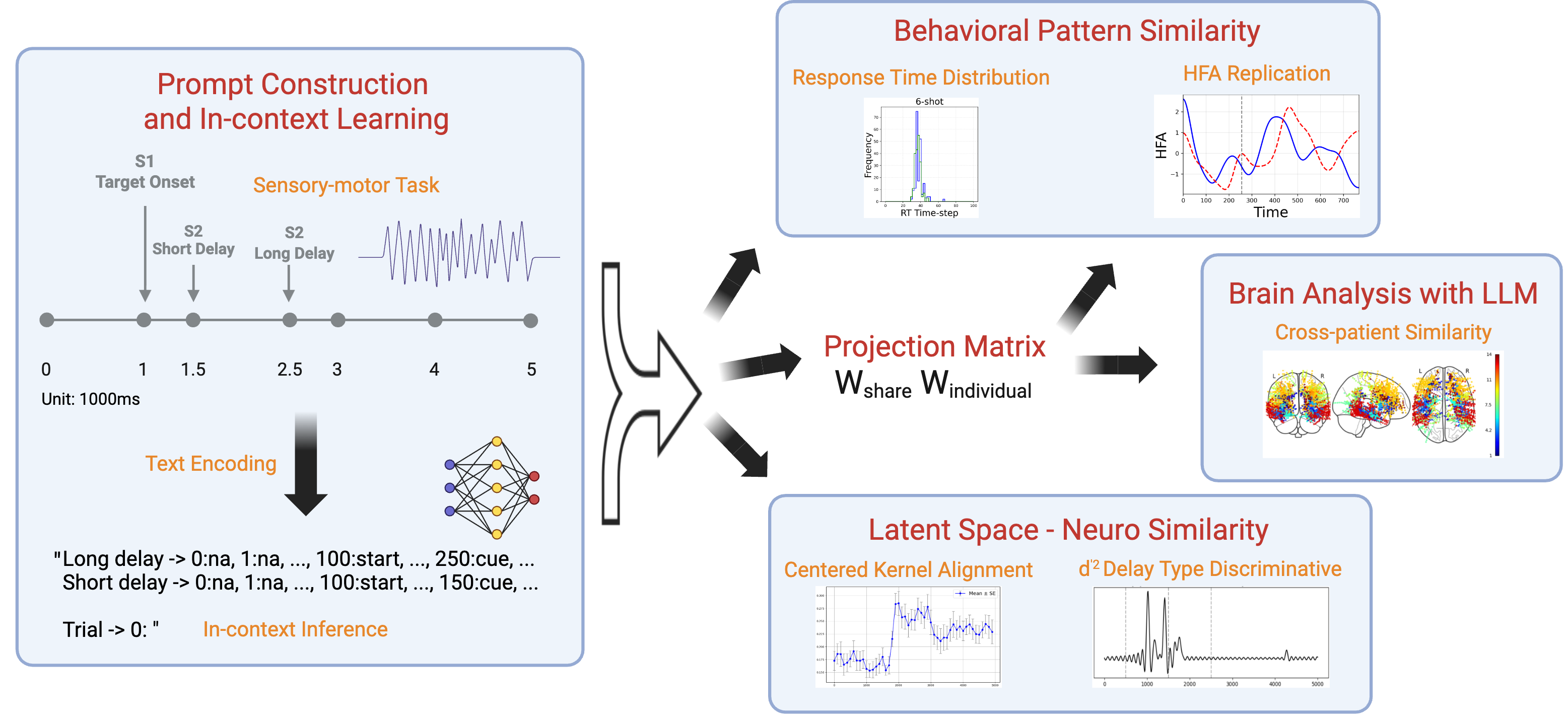}
        \caption{Experimental design and analysis pipeline for comparing human iEEG and LLM representations in a sensory-motor task. Participants performed sequential S1-S2 trials with variable delays while iEEG was recorded. An LLM processed analogous inputs via in-context inference. Analyses included response time distributions, iEEG-to-HFA transformation, projection matrix decomposition into share/individual components, $d'^2$ discriminability, and Centered Kernel Alignment for representational similarity. Cross-patient similarity maps visualized spatial alignment of neural and model representations.}

        \label{fig:overview}
    \end{figure*}

    To investigate this, we translate an established neuroscience paradigm involving anticipatory sensory-motor behavior into a task interpretable by LLMs. We translate this protocol into natural language prompts that preserve the temporal structure and sensory logic of the original task as shown in Figure \ref{fig:overview}. 
    After implementing this experiment framework on LLM, we investigated the level of alignment between LLM and human brain activities through a projection model and further similarity analysis. We see that LLM and human reaction time shares similar distribution patterns. By quantitatively measuring similarity using Central Kernel Alignment (CKA). Any level of alignment suggests that LLMs may provide insight into non-linguistic aspects of cognition and offer a scalable proxy for neuroscientist to model human cognitive activities \cite{alignment4,annual_review}. The discovery of divergences between LLM and human brain activities may also offer insights for designing human-inspired AI \cite{zhou2025divergenceslanguagemodelshuman} and for understanding the shared vs. distinct patterns in the development of artificial vs. biological intelligence. 
    
    Lastly, by aligning a non-linguistic neuroscience experiment with LLM-derived representations through the projection model, this study provides new tools for neuroscientist to study questions that are otherwise challenging to tackle. For example, in neuroscience, the problem of differentiating individual vs. shared characteristics among subjects has been a formidable topic due to the difficulty in acquiring experiment subjects, and this difficulty is further exacerbated by the fact that each subject has differing numbers and locations of electrodes. By adopting the projection model, we obtain a k-dimensional representation vector aggregated across all patients for each electrodes, which offers a new tool for neuroscientist to study the shared characteristics of brain regions in neuroscience tasks. This opens up new directions for using large-scale language models to probe foundational mechanisms of intelligent behavior across sensory, motor, and temporal domains.

\section{Methodology}
    \subsection{Sensory-Motor Task Experiment Set-up.}
        Participants performed a visual stimulus-detection task with a variable foreperiod delay. Each trial began with the onset of a fixation target (S1), followed after a short (500ms) or long (1500ms) delay by a color change cue (S2). Participants responded via button press, and response time (RT) was measured from S2 to the response. In total, we analyzed data from intraparenchymal depth electrodes distributed widely throughout the brain in all patients (mean=110.3 electrodes per participant). Intracranial EEG (iEEG) data was recorded to capture neural activity, with preprocessing following the approach in \cite{ramayya2025human} to extract high-frequency activity (HFA).

    \subsection{Text-based Temporal Alignment for LLM.}
        Each trial in the stimulus-detection task spans from S1 (fixation onset) to the end of the valid response window. The total duration of the experiment is 5 seconds. This time window is discretized into 10 ms intervals. At each time step, a keyword token represents the trial's true state, paired with its time index (starting from 0). The token mappings are as follows: \texttt{start} for S1, \texttt{cue} for S2, \texttt{button} for RT, and \texttt{na} for all other time steps. A short-delay trial with an RT at 345 ms post-S2 would be encoded as:

        \begin{framed}
        \noindent Short-delay $\rightarrow$ 0:start, 1:na, ..., 49:na, 50:cue, 51:na, ..., 83:na, 84:button, 85:na, ..., 150:na,151:na, ...
        \end{framed}
    
    \subsection{In-context Learning for Aligning LLM with Human Anticipatory Processes.}
        To model the anticipatory behavior exhibited by human participants, we leverage in-context learning using n-shot prompting, wherein the LLM is given a few example trials formatted as natural language before being asked to respond to a new, unseen instance. This approach allows the model to adapt dynamically to the structure of the task without any gradient updates or prior fine-tuning. In-context learning is particularly well-suited for replicating the experimental conditions of the original human study, where subjects were not pre-trained on the task, were unaware of the specific delay types, and had to infer temporal structure on the fly based on limited experience. This mirrors the LLM’s setup: it must infer patterns and react in real-time based on a small number of contextual examples, without knowing how the task varies across trials.

        This choice also avoids the pitfalls of fine-tuning, which risks introducing strong biases into the model’s behavior by pre-exposing it to the expected experimental structure. Our goal is not to teach the model the correct anticipatory response, but rather to assess the degree to which it can generalize or align to human-like patterns when exposed only to minimal context, which is a defining feature that signifies the superiority of LLMs as an artificial intelligence model.

    \subsection{LLM's Hidden States Projection to HFA.}
        To map the internal states of the LLM to iEEG data, we use two separate transformations: $W_{\text{shared}}$ and $W_{\text{individual}}$. This factorization is motivated by several factors. First, the approach aims to project the high-dimensional embedding space into a lower-rank shared representation space, which is subsequently mapped to the participant-specific neural activity layer. This design is supported by prior research suggesting that neural activity patterns exhibit shared characteristics across individuals, particularly under similar conditions \cite{chen2017shared,neuralsyncrony}. Consequently, $W_{\text{shared}}$ captures these shared features of brain activity. Secondly, incorporating the separate matrix $W_{\text{individual}}$ is critical due to the sparsity of iEEG data and variability in electrode configurations across participants. Given each participant may have a unique electrode setup or individual-specific neural activity patterns, $W_{\text{individual}}$ accounts for these individual differences. This structure enables the model to balance shared and unique aspects of neural activity, identifying both generalizable features and participant-specific signals. The model is expressed as:
        $$Y = X W_{\text{shared}}W_\text{individual}$$
        In this setup, $Y$ represents the true iEEG with dimensions [timesteps $\times$ electrodes], and $X$ refers to LLM's internal hidden states with dimensions [timesteps $\times$ model dimension]. $W_\text{shared}$ is modeled using an MLP with hidden dimensions [512, 128, 32, 16], and $W_\text{individual}$ is a participant-specific matrix with dimensions [16 $\times$ electrodes], where the number of electrodes varies by participant. The model is trained using \textit{Adam} optimizer \cite{diederik2014adam} with a learning rate of $1e^{-3}$ and a Mean Squared Error (MSE) loss objective for 50 epochs. No regularization is applied.
        
    \subsection{Central Kernel Alignment Analysis.}
        To quantify the similarity between LLM representations and human brain activity, we used Centered Kernel Alignment (CKA) \cite{cka}, a technique designed to measure the structural alignment of internal representations across systems. CKA is particularly well-suited for our setting, as it is invariant to isotropic scaling and orthogonal transformations, making it robust to differences in signal magnitude, dimensionality, and measurement space between human iEEG signals and LLM hidden states. Unlike simple correlation or decoding-based approaches, CKA captures relational structure within each representational space or how activity patterns across trials relate to one another, and compares these structures across systems.

        We applied linear CKA to compare hidden states from the LLM at key decision points with the corresponding iEEG responses from each participant. Linear CKA computes similarity between two feature matrices $X\in R^{n\times p_1}$ and $Y\in R^{n\times p_2}$ using their Gram (inner-product) matrices:
        $$\text{CKA}(X,Y)=\frac{||Y^\intercal{X}||^2_F}{||X^\intercal {X}||_F ||Y^\intercal{Y}||_F}$$
        This formulation captures how similarly each system encodes relationships across trials while being robust to scaling and rotation. As a sanity check, we included comparisons against a noise-matched control model to confirm that observed similarities were not due to trivial alignment or trial structure alone. High CKA values indicate that the LLM encodes a task-relevant structure that systematically mirrors the neural encoding across trials, offering a principled way to test for convergence in task-specific dynamics without assuming shared encoding formats.

    \subsection{Short vs. Long Delay Trial Discriminative Analysis.}
        Another question to ask is whether human neural activities are different when processing short-delay vs. long-delay trials, and whether LLM shows similar discriminative behavior. \cite{ramayya2025human} has identified two behaviorally and neurally distinct processes that govern human anticipatory effects on sensory-motor behavior. To characterize the extent to which humans responded differentially to the two types of trials, we calculate the statistic $d'^2$ on the representation vectors at each time point. $d'^2$ is widely adopted in analysis of neural representations \cite{d'2,RePEc:nat:nature}. Similar to max-margin, it estimates the maximum separation between the means of the representations under two types of trials. Therefore, the higher the value of $d'^2$ is, the more different short trial representations are from long trial representations. The exact expression of $d'^2$ is given by:
        $$d'^2= (\mu_1-\mu_2)^T \Sigma^{-1} (\mu_1 -\mu_2)$$
        where $\mu_1$, $\mu_2$ are the sample mean vector for short and long delay trial representations respectively.
        $\Sigma = \frac{1}{2}\left( \Sigma_1 +\Sigma_2\right)$ is the average of the corresponding sample covariance matrix for long and short delay trials. We also calculate the same $d'^2$ on projected LLM hidden states ($XW_{\text{shared}}$) at each time point to compare human and projected LLM $d'^2$ temporal patterns. We chose to conduct the analysis on the shared representation space in order to reduce the dimensionality and noise in the representations.

\section{Results}

\subsection{LLM and Human Behavior Patten Similarity. }
    \subsubsection{Raw Response Time Text Distribution}

        \begin{table}[h]
        \caption{$n$-shot Root Mean Square Error on Response Time Prediction}
        \centering
        \begin{tabular}{c|c|c|c|c|c|c}
        
        \hline
        \hline
         $n$-shot & 1 & 2 & 3 & 4 & 5 & 6 \\
        \hline
        RMSE & 27.23 & 10.12 & 8.73 & 6.11 & 7.78 & 6.38 \\
        \hline
        \hline
        \end{tabular}
        \label{table:RT_RMSE}
        \end{table}
        

        Using the text outputs from the LLM, we constructed a histogram to visualize the distribution of predicted time steps for the keyword ``\texttt{button}''. This analysis helps assess how well the model aligns with participants' true response times (RT).

        To quantify prediction accuracy across different $n$-shot settings, we computed the Root Mean Squared Error (RMSE) between the predicted and true response times \cite{hyndman2006another}. $ RMSE= \sqrt{\frac{1}{N} \sum_{i=1}^{N} (y_i - \hat{y}_i)^2}$, where $y_i$ represents the true timestep, $\hat{y}_i$ is the predicted time-step, and $N$ is the number of inferences for $n$-shot. We have $N=200$ for all $n$.
        Table~\ref{table:RT_RMSE} illustrates the average RMSE values for different $n$-shot settings under two delay conditions: short delay (500) and long delay (1500). There is a clear trend in the reduction of RMSE. LLMs' ability to learn the patients' response behavior significantly increases as we provide more examples.

        We chose $n$ with the lowest RMSE $n=6$ for the rest of the paper. The in-context learning ability of the model is reflected in how closely its predicted response time (RT) distribution approximates that of human participants. As shown in table~\ref{table:RT_RMSE}, the predicted RT distribution increasingly resembles the true distribution as the number of in-context examples $n$ increases. 

        \begin{figure}[h]
            \centering
            \includegraphics[width=0.45\textwidth]{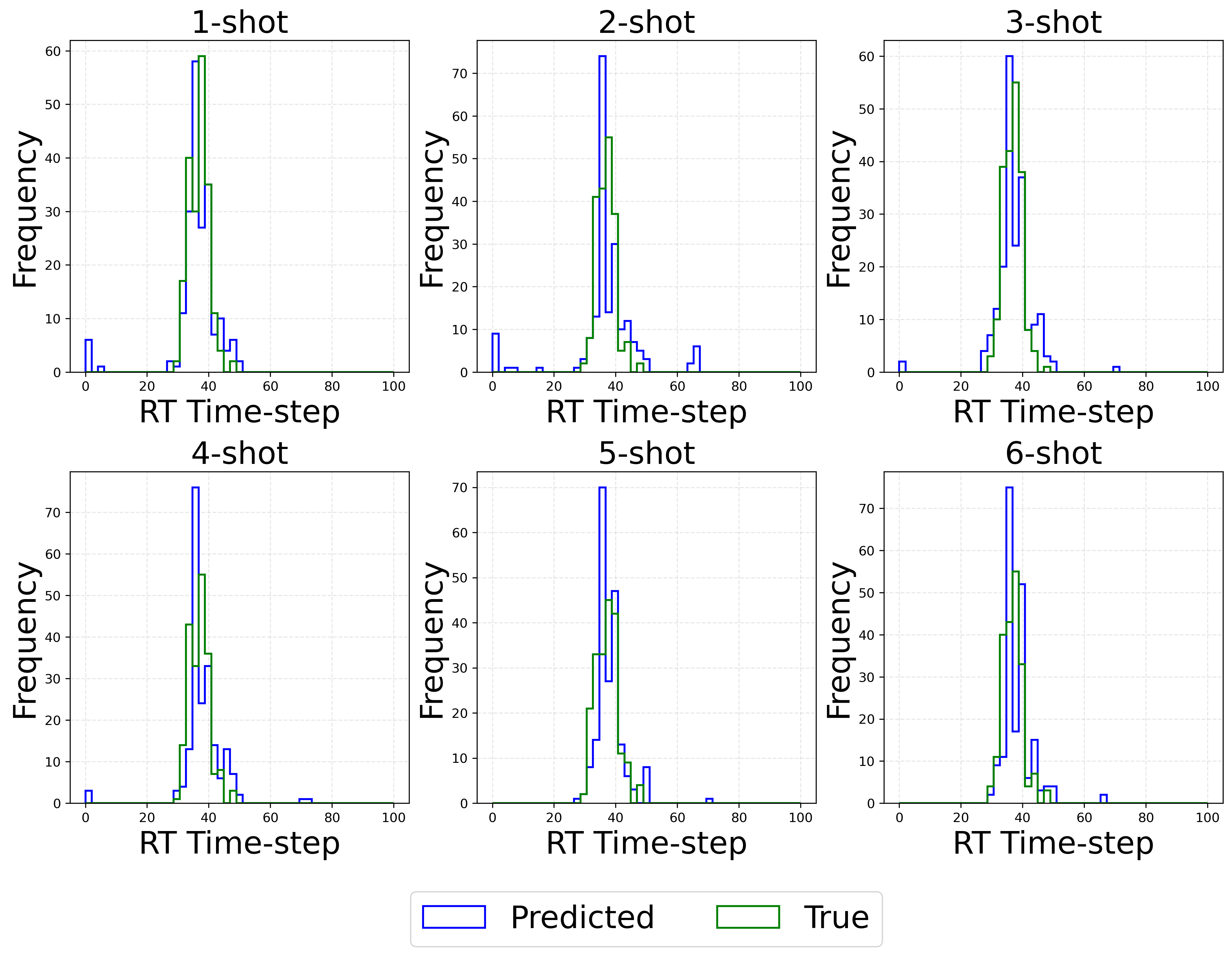}
            \caption{Distribution of predicted response times (RT) compared to true RTs for $n$-shot in-context learning. LLM responds pattern closely mimic human pattern.}
            \label{fig:rt_distribution}
        \end{figure}

    \subsubsection{Projection Model}
        Training the projection matrices $W_{\text{shared}}$ and $W_\text{individual}$ resulted in concurrent loss reduction on both training and evaluation sets. We first qualitatively examine how well the LLM-derived features align with participants' neural activity.
        \begin{figure}[h]
            \centering
        \includegraphics[width=0.4\textwidth]{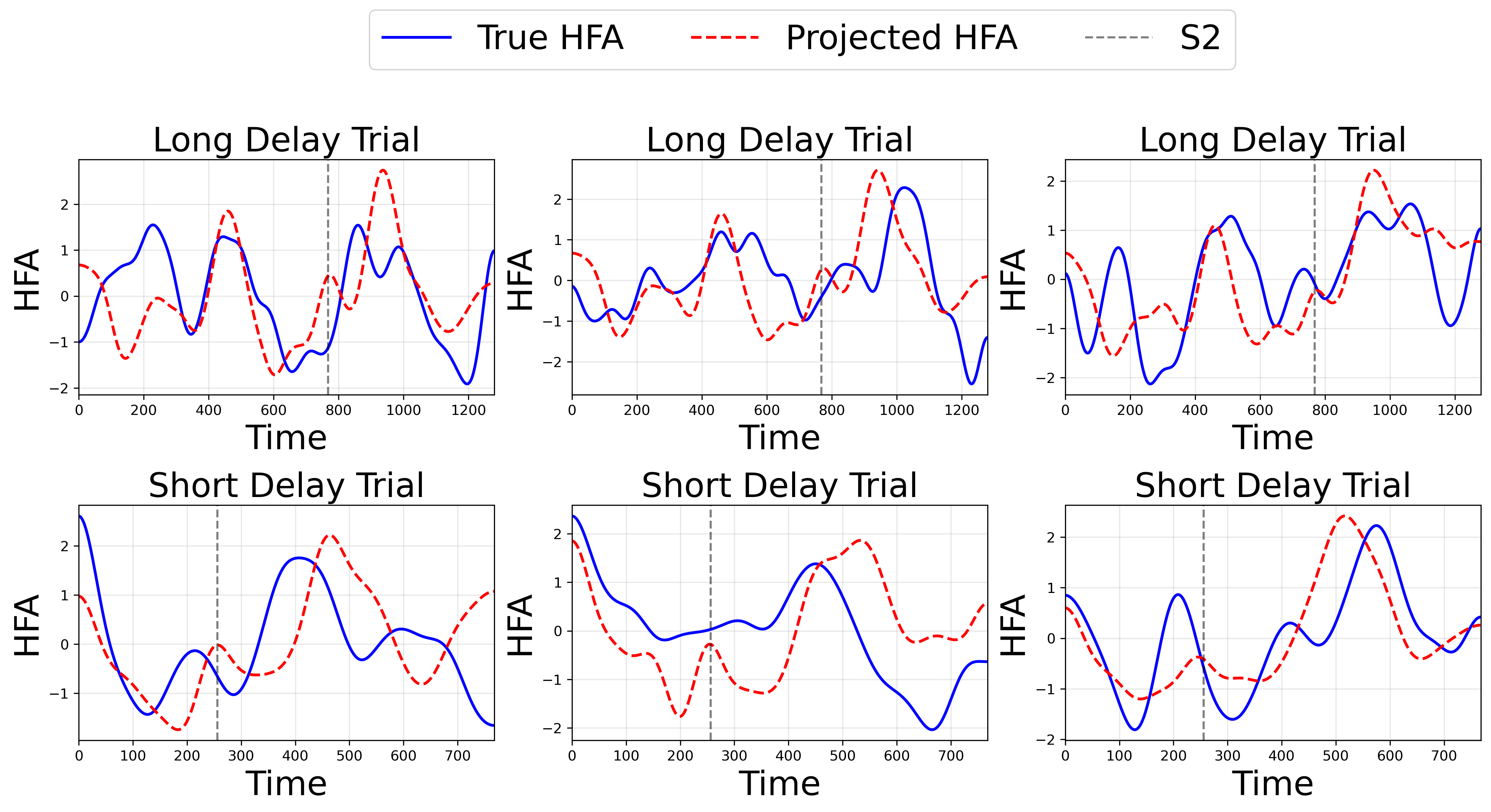}
            \caption{Comparison of true HFA power (blue) 
            and projected HFA (red) across multiple trials.}
            \label{fig:projected_hfa}
        \end{figure}
        
        As shown in Figure~\ref{fig:projected_hfa}, the projected HFA closely follows the temporal dynamics of the true signal across both delay conditions. The projected HFA (Figure ~\ref{fig:projected_hfa}) confirms that the learned projection effectively maps the LLM’s hidden states to participants’ neural responses. While the projected HFA does not perfectly align with the true signals, it successfully captures underlying temporal relationships. This suggests that, despite the complexity of both systems, meaningful correspondences can be identified through a learnable mapping. Although neural activity and LLM-derived representations are shaped by distinct underlying mechanisms, their shared temporal structures suggest a potential link between cognitive processing and model-inferred representations. The ability of a learnable projection to reveal such alignment supports the hypothesis that there exists the same structured changes in both biological and artificial systems. These findings highlight the potential for further investigation into how language models can approximate certain patterns of human neural dynamics. This suggests that LLM representations can be mapped onto neural activity through a learned linear projection, which suggests a potential link between cognitive processing and model-inferred representations. We later quantify this alignment using Centered Kernel Alignment (CKA) to assess representational similarity.

    \subsection{ LLM and Humans Latent Space Similarity}
        \subsubsection{Short Delay vs. Long Delay Trial Discriminative Analysis.}
        
\begin{figure}[h]
    
        \begin{minipage}{0.23\textwidth}
            \includegraphics[width=\textwidth]{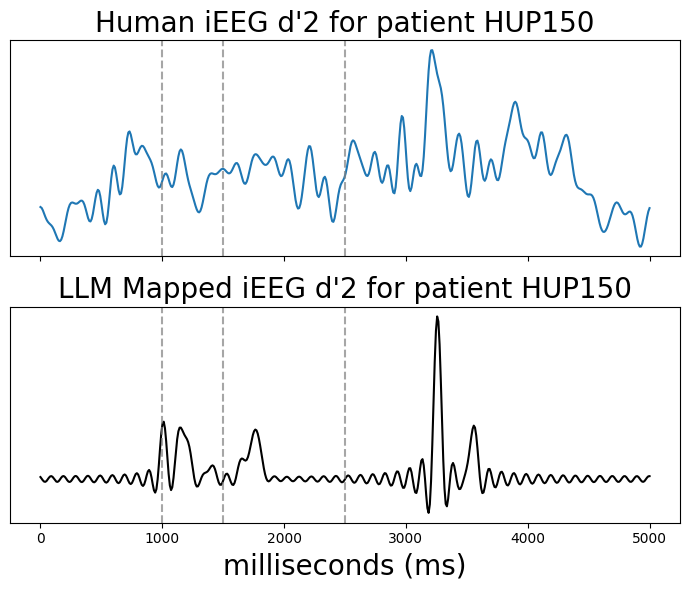}
        \end{minipage}
        \begin{minipage}{0.23\textwidth}
            \includegraphics[width=\textwidth]{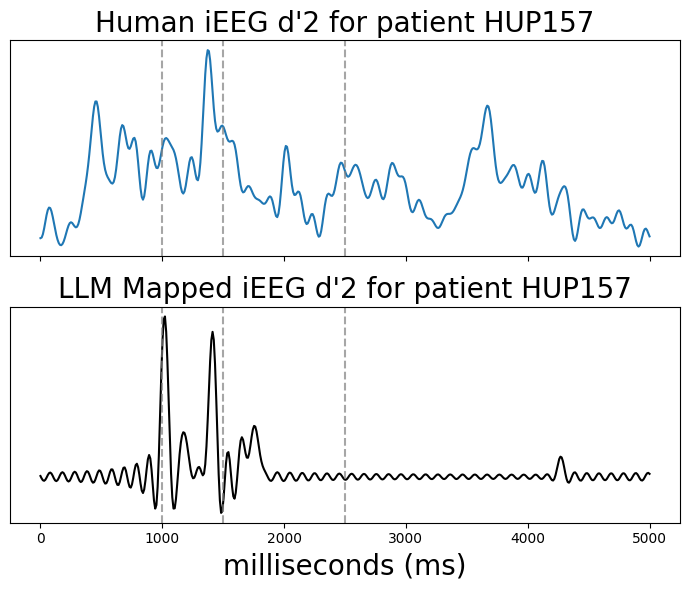}
        \end{minipage}
        \begin{minipage}{0.23\textwidth}
            \includegraphics[width=\textwidth]{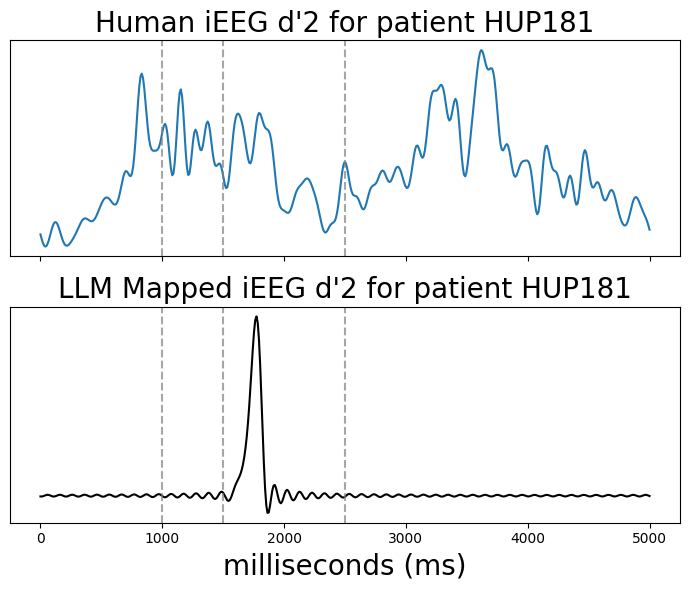}
        \end{minipage}
        \begin{minipage}{0.23\textwidth}
            \includegraphics[width=\textwidth]{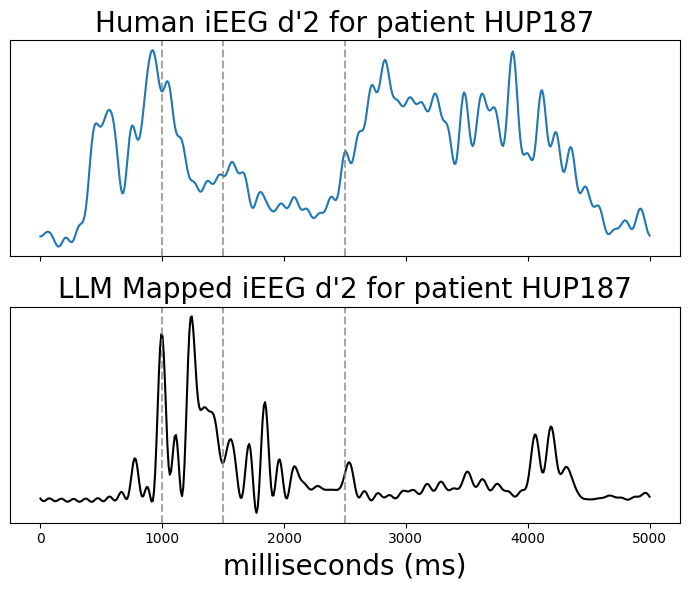}
        \end{minipage}
        \caption{Time Series of $d^{'2}$ of Human iEEG vs $d^{'2}$ of Projected LLM. Higher value of $d'^2$ indicates representations in short and long trials are more different at that time point. }
        \label{d2_comparison}
           
        \end{figure}

         Figure \ref{d2_comparison} shows examples of the comparisons during the experiment interval [0 ms, 5000 ms]. Recall that a higher value of $d'^2$ signifies a higher level of differentiation of responses between short and long trials, therefore we focus on the location of peaks of $d'^2$. The differentiability peaks are located at comparable time points for humans and projected LLM hidden states. As S1 begins at 1000 ms, we observe some jumps in both human and LLM $d'^2$. The short delay signal (S2) starts at 1500 ms, and we can see peaks appearing around the time frame [1000 ms, 2000 ms] in the LLM hidden states representations, especially around 1500 ms. The fact that the peaks appearing before the actual onset at 1500 ms indicates LLM also captures some aspects of human anticipation. This phenomenon can be seen in patients HUP150, HUP157, HUP187 in Figure \ref{d2_comparison}. To quantitatively compare the similarity of the temporal pattern of  $d'^2$, we divided the time frame into intervals as indicated by the vertical lines in Figure \ref{d2_comparison}, and then we compared how close the locations of the peaks of $d'^2$ are for humans and LLM within each interval. We calculated the proportions of samples whose LLM and human maximum $d'^2$ locations differ within a certain threshold (50ms, 150ms 250ms.) in Table \ref{d2_table}. The overall alignment of the peak locations suggests the projected LLM states exhibit similar discriminative structure against short and long trials. Without further training LLMs on human reaction patterns, LLMs can capture the subtle reaction difference between short and long trials with only in-context learning, signifying potential alignment between artificial intelligence systems and human nervous systems.
        
        \begin{table}[h]
        \caption{Proportions of samples whose LLM and human peak $d'^2$ locations differ within a certain threshold under various time intervals}
        \centering
        \begin{tabular}{c|c|c|c}
        
        \hline
        \hline
         & 1000-1500ms & 1500-2500ms & 2500-3500ms \\
        \hline
        50ms & 0.29 & 0.14 & 0.07 \\
        \hline
        150ms & 0.71 & 0.21 & 0.14  \\
        \hline
        250ms & 0.85 & 0.50 & 0.42\\
        \hline
        \hline
        \end{tabular}
        \label{d2_table}
        \end{table}
       
    \subsubsection{Neural Representational Alignment}
        To assess the alignment between LLM representations and human neural activity, we computed Centered Kernel Alignment (CKA) in two settings:
        \subsubsection{Global Alignment}
            We computed CKA scores using the full time series of LLM hidden states and iEEG features, and compared them against each other as well as against random noise baselines. As shown in Table~\ref{table:cka-3-way}, LLM and human iEEG data exhibit moderate alignment (CKA = 0.3926), while their similarity with noise is negligible—validating that the shared structure is non-trivial.
            
            \begin{table}[h]
                \caption{Three-way CKA similarity scores between Human, LLM, and Noise representations}
                \label{table:cka-3-way}
                \centering
                \begin{tabular}{c|c|c|c}
                    \hline\hline
                    -- & Human & LLM & Noise \\
                    \hline
                    Human & 1.0000 & 0.3926 & 0.0734 \\
                    LLM   & --     & 1.0000 & 0.0174 \\
                    Noise & --     & --     & 1.0000 \\
                    \hline\hline
                \end{tabular}
            \end{table}
        
        \subsubsection{Temporal Dynamics Alignment}
\begin{figure}[h]
                \centering
                \includegraphics[width=0.45\textwidth]{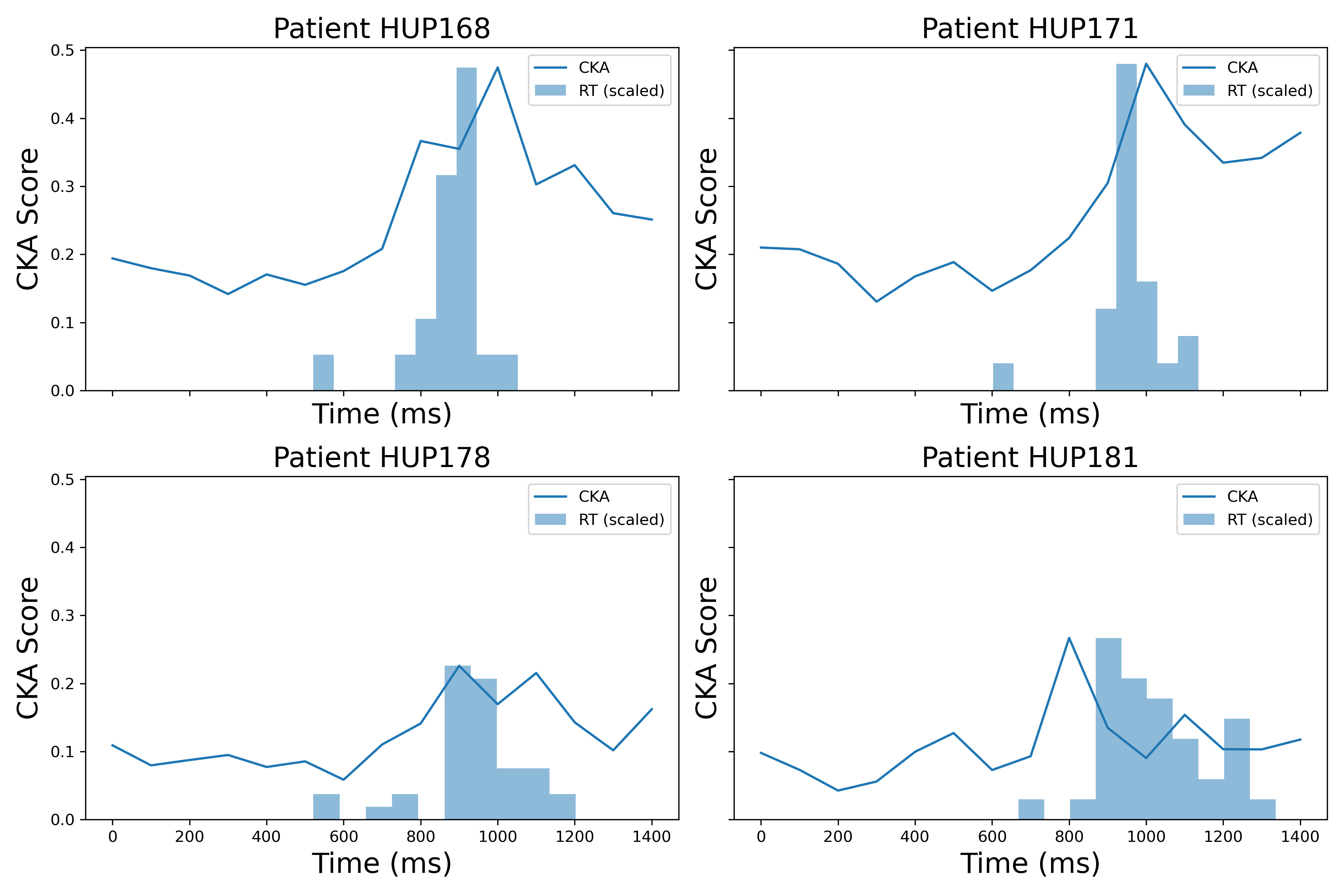}
                \caption{RT distribution vs. LLM-Human CKA Score over 15 time bins.}
                \label{fig:cka/rt_dist_cka}

                \centering
                \includegraphics[width=0.45\textwidth]{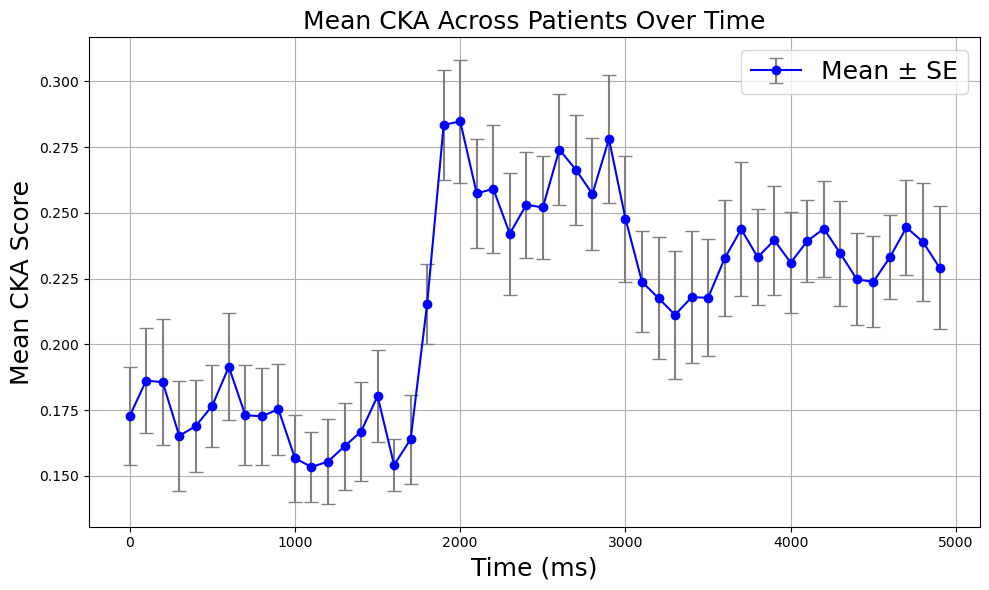}
                \caption{LLM-Human Mean CKA over 15 time bins.}
                \label{fig:cka/mean_cka_with_se}
            \end{figure}

            To explore how representational similarity evolves over time, we divided each short-delay trial into 15 consecutive time bins and computed CKA between LLM and iEEG representations within each bin. The resulting temporal profile (Figure~\ref{fig:cka/rt_dist_cka}) reveals that alignment increases significantly around the time of the behavioral response. This behavior can be seen across all patients (Figure~\ref{fig:cka/mean_cka_with_se}).

            Timestep with peak CKA scores align closely with time bins showing heightened brain activity. The distribution also corresponds with the that of response times across participants. One interpretation is that prior to S2, both LLM and neural signals are less structured, resulting in low similarity. When the stimulus appears and participants respond, both systems exhibit more structured activity, increasing representational alignment.

    \subsection{Understanding Human Brain via LLMs}
    As the projection model provides $W_{\text{individual}}$ that encodes a k-dimensional vector for each of the electrodes, we naturally have a feature for each of the electrodes across all patients. We measured the average cosine similarity among electrodes grouped by the different areas of interests (ROI). As a control, we also calculated the average cosine similarity of $N$ independent k-dimensional gaussian random vectors, where $N$ corresponds to the sample size of electrodes belonging to each ROI. We calculated the metric $p$ by sampling $N$ k-dimensional gaussian random vectors 1000 times and each time calculates their average cosine similarity, and $p$ represents the proportion of the times that the average cosine similarity of the k-dimensional gaussian vectors exceeds that of the k-dimensional feature vectors of the electrodes. Lower values of $p$ indicates more evidence for higher than random similarity among electrodes in the specific ROI.
        \begin{figure}[h]
                \includegraphics[width=0.45\textwidth]{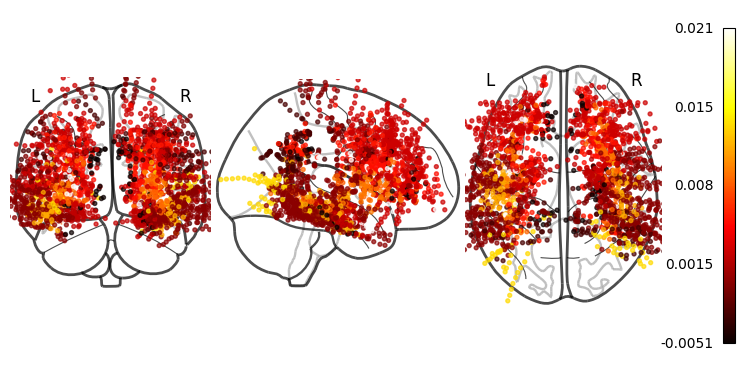}
                \caption{Average Cosine Similarity of Electrodes by ROIs (Calculated based on LLM Derived Representations)} \label{individual_brain}
        \end{figure}
    
    We can see in Table \ref{table_roi} that the ROIs with low values of $p$ are: SLF WM, Prefrontal, Arc/Unc Fasiculus, Insula, and ILF-MLF WM. This indicates that, across all patients, electrodes in these regions exhibit higher than random similarity during the sensory-motor experiment.
    
    With further analysis, we found that these regions are responsible for the sensory and motor-related neural responses. The SLF WM area is known for governing attention \cite{SLF}, a function relevant in this experiment. The prefrontal cortex is also known to receive input from multiple cortical regions to process "in the moment" information \cite{prefrontal}. The Arc/Unc Fasiculus region is also known to be responsible for visuospatial processing among other tasks \cite{AF,AF2}. The ILF-MLF WM region, which includes the medial longitudinal fasciculus (MLF), is known to control muscle movement for horizontal eye movement, allowing for a change in the fixation point \cite{MLF}. The inferior longitudinal fasciculus (ILF) is speculated to be associated with fast transfer of visual information \cite{ILF1,ILF2}, which is also relevant in the setup of the experiment where subjects are required to look at visual signals. This result supports the findings in neuroscience research, and it demonstrates the potential for using LLMs to better understand human brain.
        \begin{table}[h!]
        \centering
        \caption{Cosine Similarity and P-values for Different Brain Regions}
        \begin{tabular}{l c c}
        \hline
        \textbf{ROI} & \textbf{Cosine Similarity} & \textbf{p} \\
        \hline
        SLF WM                & 0.0054  & \textbf{0.03} \\
        Perirolandic-CST      & -0.0005 & 0.86 \\
        Prefrontal            & 0.0025  & \textbf{0.00} \\
        Arc/Unc Fasiculus     & 0.0212  & \textbf{0.00}   \\
        Insula                & 0.0092  & \textbf{0.02} \\
        Parietal              & -0.0027 & 0.33 \\
        Cingulate             & -0.0051 & 0.27 \\
        ILF-MLF WM            & 0.0111  & \textbf{0.00} \\
        Thalamocortical WM    & 0.0050  & 0.04 \\
        MTL                   & 0.0020  & 0.44 \\
        Temporal Lobe         & -0.0001 & 0.90 \\
        Occipital             & 0.0131  & \textbf{0.01} \\
        IFOF WM               & 0.0085  & 0.04 \\
        Striatum              & 0.0081  & 0.47 \\
        \hline
        \end{tabular}
        \label{table_roi}
        \end{table}
    
\section{Related Work}
    Several works have also studied the alignment between human brain activities and LLM internal states from other perspectives. The authors in \cite{alignment1} show that artificial intelligence models trained to predict masked words from a large amount of text can generate activations similar to those of the human brain. \cite{alignment2} studied the next token prediction behavior in autoregressive deep language models and assessed their ability to linearly map onto the human fMRI and MEG. Besides a computational view, another perspective of human and LLM similarity examines the shared geometric structure \cite{alignment3}, where the authors show similarity in the embedding space of IFG region of the brain and the latent space of deep language models in a audio podcast processing task. Following similar idea, We expand upon these works by going beyond language based tasks and demonstrate that the alignment extends beyond languages.

\section{Conclusion}
    Our findings demonstrate that the LLM is able to replicate patient-specific behavioral patterns, particularly the distribution of response times across different delay conditions. Moreover, this behavioral alignment was accompanied by neural alignment. Using a linear projection method, we found that the LLM's internal states could approximate high-frequency activity (HFA) in the human brain during the decision period, particularly in the moments following the second stimulus (S2), where anticipatory and motor planning processes are most active.
    
    In all, these results highlight the existence of a potential pathway between language-trained model representations and human neural dynamics even in domains that extend beyond language itself. Although we do not claim that LLMs perform sensory-motor processing in a human-like fashion, the structured alignment we observe suggests that there might exist a pathway between the representational geometry of LLMs and the human brain system when engaged in predictive and temporally structured tasks.
    
    The pathway between LLM and human brain activity is represented by $W_{\text{share}}$ and $W_{\text{individual}}$. Using these two quantities, we provided tools for neuroscientist to analyze cross-subjects similarity in brain activities. This work suggests a broader framework for using language models as computational tools for cognitive neuroscience: not only to simulate language processing, but also to probe cognitive functions such as anticipation, and sensorimotor integration. In doing so, we create a promising path for a bidirectional research potential, where insights from human cognition can inform AI model development, and emerging behaviors in LLMs can, in turn, guide neuroscience toward new hypotheses about the structure and function of intelligent behavior.

\bibliography{aaai2026}

\end{document}